\documentclass[12pt,preprint]{aastex}
\usepackage{graphicx,natbib}
\usepackage{emulateapj5}
                                                   
\begin{document} % ----------------------------------------------
                               
% Title and Authors ----------------------------------------------
\title{The Outskrits of Spiral Galaxies: Evidence for Multiple 
       Stellar Populations}
                                            
\author{M. Mouhcine\altaffilmark{1,2}}
\altaffiltext{1}{Astrophysics Research Institute, Liverpool John 
Moores University, Twelve Quays House, Egerton Wharf, Birkenhead, 
CH41 1LD, UK. }
\altaffiltext{2}{Isaac Roberts Fellow}
                                                         
% Abstract -----------------------------------------------------
\begin{abstract}

We present an analysis of the metallicity distribution functions 
of fields projected along the minor axis for a sample of inclined 
spiral galaxies in order to search for evidence of the presence of 
multiple stellar populations. In all cases, the stellar populations 
appear to have asymmetric metallicity distributions with very high 
confidence levels. The mean metallicities of both stellar 
subpopulations, determined from mixture modelling of the metallicity 
distribution functions, correlate with parent galaxy luminosity. 
This suggests that the vast majority of field stars have probably 
formed in galactic fragments that were already embedded in the dark 
matter halo of the final galaxy. The steeper correlation between 
the mean stellar metallicity and parent galaxy luminosity is driven 
by an increasing fraction of metal-rich stars with increasing galaxy 
luminosity. Metal-poor components show larger dispersion in 
metallicity than metal-rich components. 
These properties are in striking similarity with those of globular 
cluster subpopulations around early-type galaxies. The properties 
of field stars along the minor axis are consistent with a formation 
scenario in which the metal-poor stars formed in all galaxies, 
possibly as a result of tidal disruption of dwarf-like objects. 
An additional metal-rich component might be related to the formation 
of the bulge and/or the disk.

\end{abstract}

% Section 1 : Introduction -------------------------------

\section{Introduction}

Stars in the outskirts of galaxies are among the oldest and the 
most metal-poor stellar components of galaxies. Therefore their 
properties are clues to the understanding of how galaxies have 
assembled their mass. 
Remarkably little is known about these populations in other 
spiral galaxies, and their formation histories. Recently, 
Mouhcine et al. (2005a,b) presented a study of the properties 
of stellar populations along the minor axis of a sample of 
highly inclined spiral galaxies beyond the Local Group. 
Their mean metallicities are found to correlate with parent 
galaxy luminosity, in the sense that bright galaxies tend to 
have more metal-rich stars than faint galaxies. 
The observed metallicity distribution functions show sharper 
cut-off at the metal-rich end, with an excess of intermediate 
metallicity stars in comparison to the simple chemical evolution 
model predictions. Interestingly, the stellar populations in 
the outskirts of early type galaxies have similar properties to 
those of spiral galaxies at a similar parent galaxy luminosity. 

One of the most noteworthy fundamental findings in galaxy property 
studies is the bimodal nature of metallicity/color distributions 
of globular cluster systems of (early-type) galaxies (e.g., Peng 
et al. 2006 in references therein). 
A large body of evidence has established that the properties of 
globular cluster systems of early-type galaxies are linked to 
those of their host galaxies (e.g., van den Bergh 1975; Brodie 
\& Huchra 1991; Forbes et al. 1997; Kundu \& Whitmore 2001, 
Larsen et al. 2001).
The globular cluster formation scenarios discussed to date 
predict all a link between metal-rich globular clusters and 
field stars (e.g., Ashman \& Zepf 1992; Forbes et al. 1997; 
Strader et al. 2004). This is supported by observation of 
several globular cluster systems (e.g., M~31: Jablonka et al. 
2000; NGC~5128: Harris \& Harris 2002; NGC 1399: Forte et al. 
2005), and the nearly constant {\it bulge specific frequency} 
in elliptical galaxies and spiral bulges (Forbes et al. 2001). 
Forte et al. (2005) have discussed the arguments to support 
the conclusion that field stars may show two major stellar 
populations that share the properties of each globular cluster 
subpopulation. For the particular case of the early-type galaxy 
NGC~1399, which exhibits a bimodal globular cluster metallicity 
distribution, the galaxy surface brightness profile, the color 
gradient and the behavior of the cumulative globular cluster 
specific frequency, are compatible with the presence of two 
dominant diffuse stellar populations, associated with blue 
and red globular cluster subpopulations (Forte et al. 2005). 
The reconstructed metallicity distribution function of field 
stars in the outskirts of NGC~1399 shows striking similarities 
with those of the outer fields of the early-type galaxy NGC~5128 
(Harris \& Harris 2000), and M~31 (Durrell et al. 2001, 2004), 
i.e., strongly weighted to moderately high metallicities with 
a metal-poor tail extending to ${\rm [Fe/H]\sim\,-2.5}$.

The observed metallicity distribution function of NGC~5128
field stars is matched reasonably well by the predicted 
distribution of the accreting box chemical evolution model, 
as long as the star formation took place over an {\it 
extended period}, simultaneously with a rapid inflow of very 
metal-poor gas that declines exponentially with time (Harris 
\& Harris 2002). However, there is {\it no} accreting box 
chemical evolution model that could reproduce the shape of 
the metallicity distribution function of the galaxy globular 
cluster system (Vandalfsen \& Harris 2004). Using the accreting 
box chemical evolution models to reproduce the shapes of field 
star metallicity distribution functions, and mixture models 
to reproduce those of globular cluster systems, ignores the 
observational evidence supporting a likely connection between 
globular cluster systems and field stars. 
Interestingly, by fitting mixture models to the metallicity 
distributions of field stars in the outskirts of M~31 and 
NGC~5128, Durrell et al. (2001) and Harris \& Harris (2000) 
have found that the metal-rich and metal-poor peak 
metallicities are similar to those of globular cluster 
systems of both galaxies. Note however that the metallicity 
distribution functions of field stars have significantly 
different shapes from those of globular cluster systems for 
both M~31 and NGC~5128. While field stellar populations are 
dominated by metal-rich stars, i.e., ${\rm [Fe/H] \sim -0.5}$, 
with a metal-poor tail, far more globular clusters are 
metal-poor (see Harris \& Harris 2002 and Forte et al. 2005 
for a possible interpretation of the observed increase of 
the number of globular clusters per unit field star numbers 
at low metallicity).

Little is known about the globular cluster systems of spirals.
Goodfrooij et al. (2003) have shown that their properties are 
consistent with a scenario in which globular cluster systems 
are made up of a {\it universal} halo population of globular 
clusters that is present around each galaxy, and a second 
population associated with the bulge, which grows approximately 
linearly with its mass (see also Chandar et al. 2004). However, 
a few spirals did not show evidence for systems of inner 
metal-rich globular clusters, consistent with a scenario of 
bulge formation through secular processes (see Olsen et al. 
2004). By comparing the properties of globular cluster systems 
for a sample of spirals and early-type galaxies, Forbes et al. 
(2001) have concluded that the formation of globular cluster 
systems of both class of galaxies might share similarities.
The observed similarities between the properties of field 
stars along the minor axis of bright spirals and those at the 
outer parts of early-type galaxies suggest that the conclusion 
of Forte et al. (2005), i.e., there is multiple diffuse stellar 
populations in the outer parts of early-type galaxies, could be 
extended to field stars in the outskirts of spirals. The aim of 
this paper is to look for further evidence for multiple stellar 
population along the minor axis using the approach used 
extensively for globular cluster systems, i.e., a fitting of 
mixture models to field star metallicity distribution functions  
constructed by Mouhcine et al. (2005c). 

The layout of this paper is as follow; in \S~\ref{data} we 
present briefly the data set. In \S~\ref{analysis} we report 
correlations between the properties of the stellar populations 
in spiral galaxy outskirts and host galaxy luminosities. 
The implications of our finding for the formation and evolution 
of the stellar populations in the outer regions of galaxies are 
discussed in \S~\ref{impli}. 

\section{Galaxy sample and metallicity distribution functions}
\label{data}

Mouhcine et al. (2005a,b,c) have observed a sample of spiral 
galaxies using the Wide Field Planetary Camera 2 on broad the 
Hubble Space Telescope with the aim of unambiguously resolving 
their stellar populations along the minor axes down to one or 
two magnitudes below the tip of the red giant branch.
The galaxy selection criteria, observations, and the data 
reduction have been described in detail by Mouhcine et al. 
(2005c). We therefore only briefly summarize this information 
here. The sample consists of nearby inclined spirals located 
at high Galactic latitude, i.e., $|b|>30^{\circ}$, with 
morphological types from Sa to Sc. The observed fields locate
well outside the visible outskirts of the disks and bulges on 
the Palomar Sky survey. Photometric errors and incompleteness 
were estimated over the color-magnitude diagram by using 
artificial star experiments. The depth of the observations vary 
from galaxy to galaxy, with 50\% completeness limits ranging 
from red giant star absolute I-band magnitudes $-0.9$ to $-3.5$.

The sensitivity of photometric properties of red giant stars 
to metallicity rather than age offers an opportunity to derive 
first order metallicity distribution functions of stellar 
populations from photometric data. To derive the metallicity 
distribution functions for the galaxy sample, the locations of 
red giant stars in the color-magnitude diagrams were compared 
to their predicted location (e.g., Harris \& Harris 2000; 
Sarajedini \& Van Duyne 2001; Brooks et al. 2004; Tiede et al. 
2004). A dense fiducial grid of theoretical red giant branch 
tracks from VandenBerg et al. (2000) on the M$_I$/(V-I) diagram 
has been superimposed on the foreground extinction-corrected 
data, and interpolate between them to derive an estimate of 
stellar metallicity on a star-by-star basis. All the models 
we used assume that the stars are $\alpha-$enhanced, i.e., 
[$\alpha$/Fe]=0.3. To cover stars located beyond the most 
metal-rich track of the theoretical grid, we add the observed 
red giant branch fiducial of the old metal-rich disk open 
cluster NGC~6791 (Taylor 2001). To avoid biasing the derived
metallicity distribution functions, stars that are one time 
the (V-I) color error bluer, at a given reddening-corrected 
I-band magnitude, than the lowest metallicity track, as well 
as for stars redder than the highest metallicity fiducial 
track, were included to construct the metallicity distribution 
functions. Their metallicities were estimated by extrapolating 
the relationship between the colors of fiducial tracks at a
given reddening-corrected I-band magnitude and their 
metallicities. This leaves a very small number of stars out 
the sample considered for building the metallicity distribution 
functions. The color-magnitude diagrams of the observed fields 
are dominated by old red giant stars, with no clear evidence 
for the presence of an intermediate age stellar population 
component, i.e., younger than $\sim 4$ Gyr (see Mouhcine et al.
2005c). The error in assuming a $\sim 12$ Gyr age instead of a 
$\sim 6$ Gyr is that the metallicity is underestimated by 
0.2-0.3 dex, of order the errors of the calibration used to 
estimate stellar metallicity from optical photometry. 

For nearly all sample galaxies, apart from NGC~4258, the $50\%$ 
completeness limit falls well below the brightest end of the red 
giant branch to allow accurate measurements of the metallicity 
distribution functions (see Fig. 2 and Fig. 3 of Mouhcine et al. 
2005c). To account for the effects of 
incompleteness, each star was counted as the inverse of the 
photometric completeness at the location of the star on the 
color-magnitude diagram. For NGC~4258, the photometry does not 
probe the red giant branch far, and the 50\% completeness level 
limit falls close to the brightest end of the red giant branch 
sequence. We have limited ourselves to a range of I-band 
magnitudes where the completeness level is not severely low, 
i.e., higher than 50\%. However, stars with metallicities higher 
than ${\rm [Fe/H]\sim\,-0.7}$ lie at fainter magnitudes than the 
magnitude range chosen to construct the metallicity distribution 
function of stars at the outskirts of NGC~4258. For this galaxy, 
the metal-rich end of the metallicity distribution function is 
suspected to be incomplete, i.e., a fraction of metal-rich stars 
are missing.

\section{Results for metallicity distribution functions}
\label{analysis}

\begin{figure*}
\includegraphics[height=2.in,width=2.in]{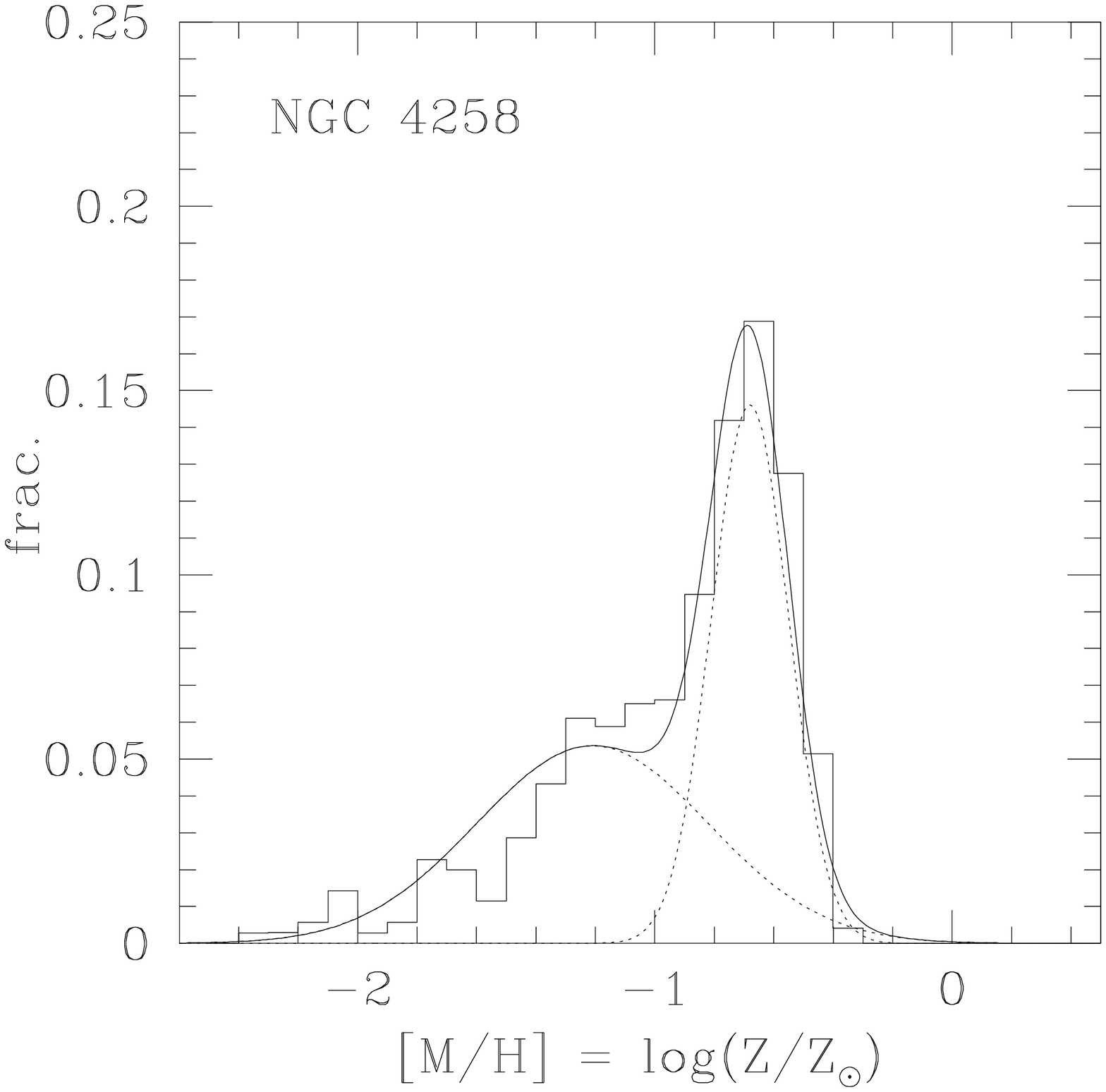}
\includegraphics[height=2.in,width=2.in]{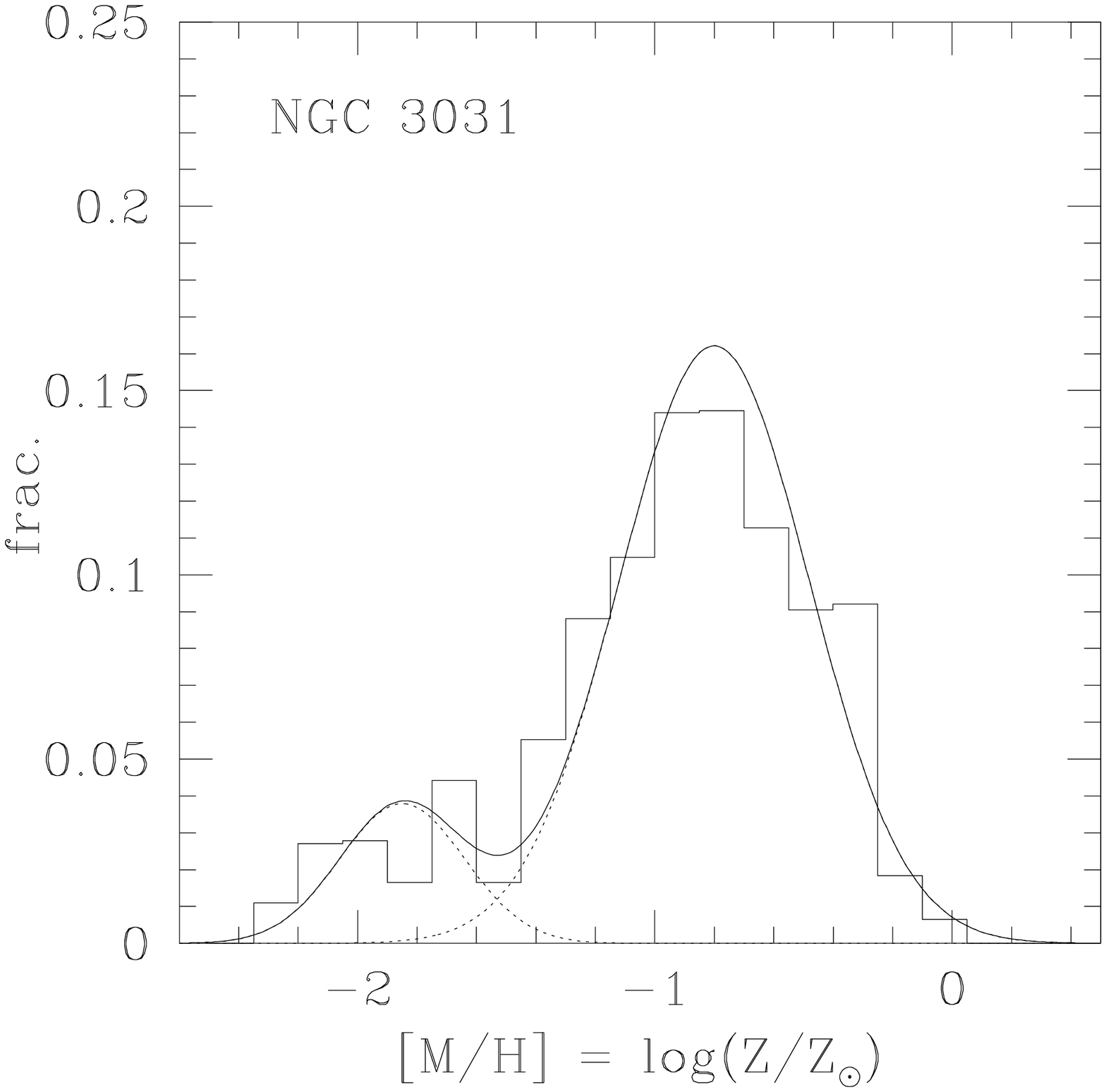}
\includegraphics[height=2.in,width=2.in]{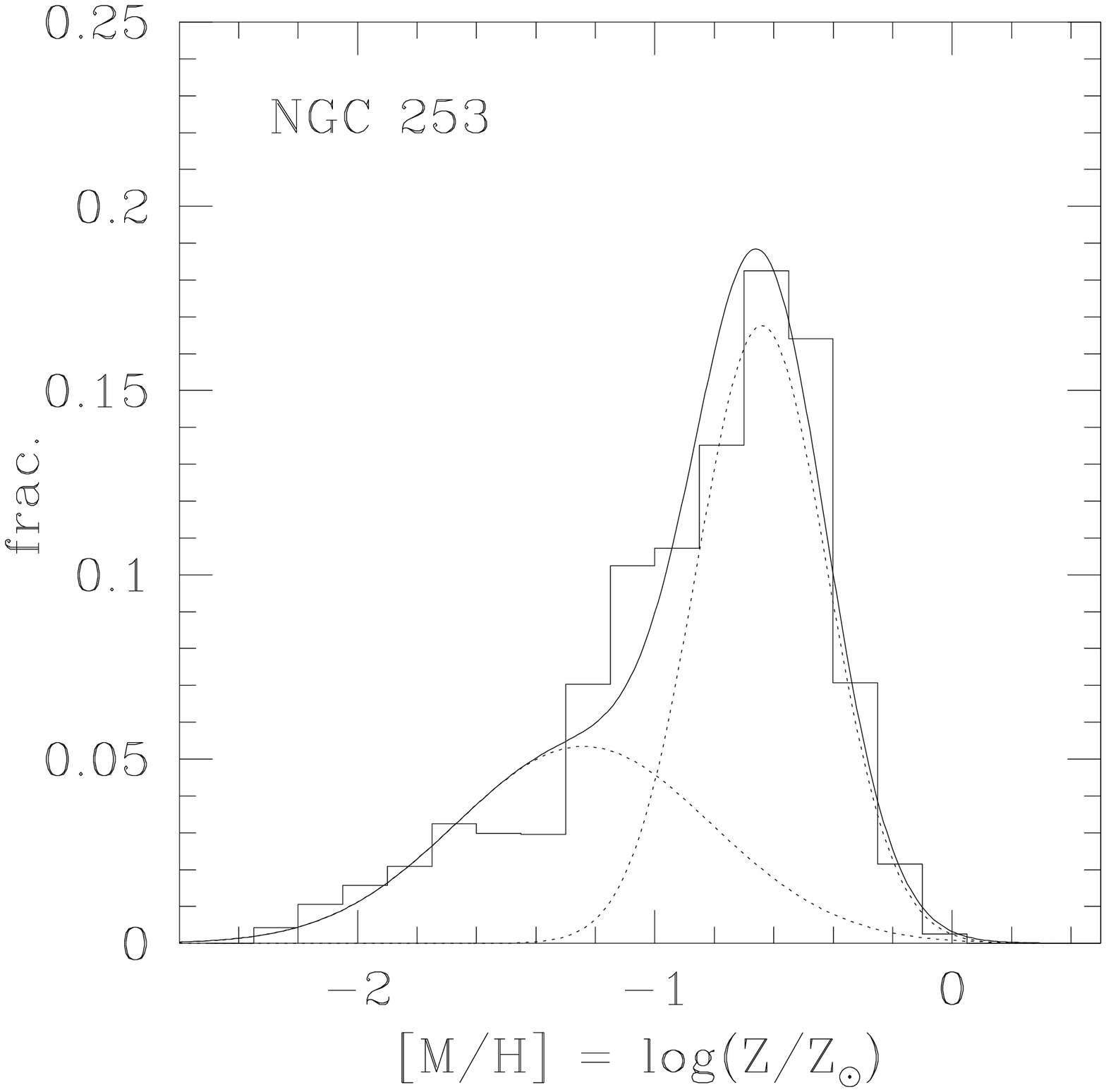}
\includegraphics[height=2.in,width=2.in]{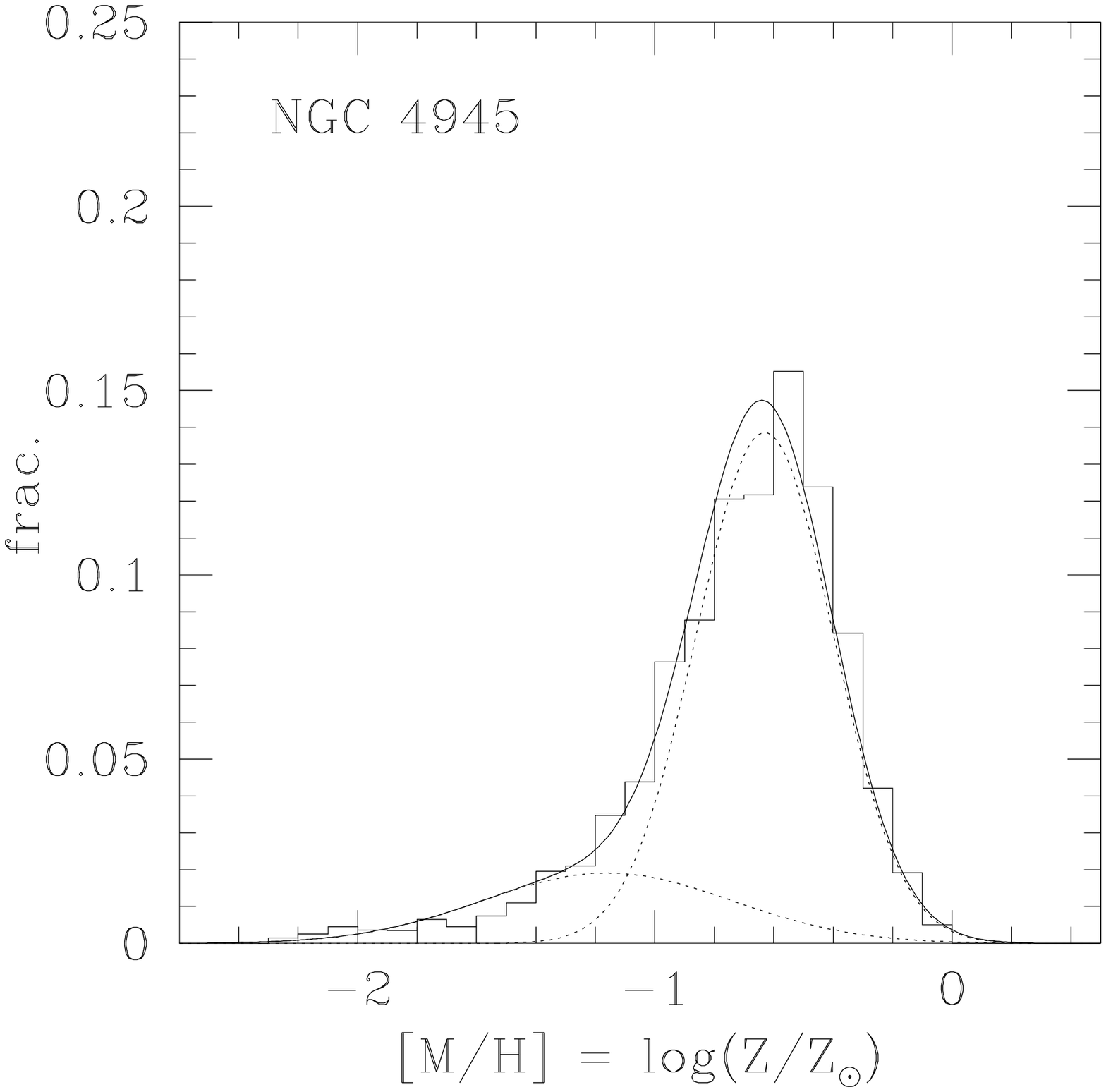}
\includegraphics[height=2.in,width=2.in]{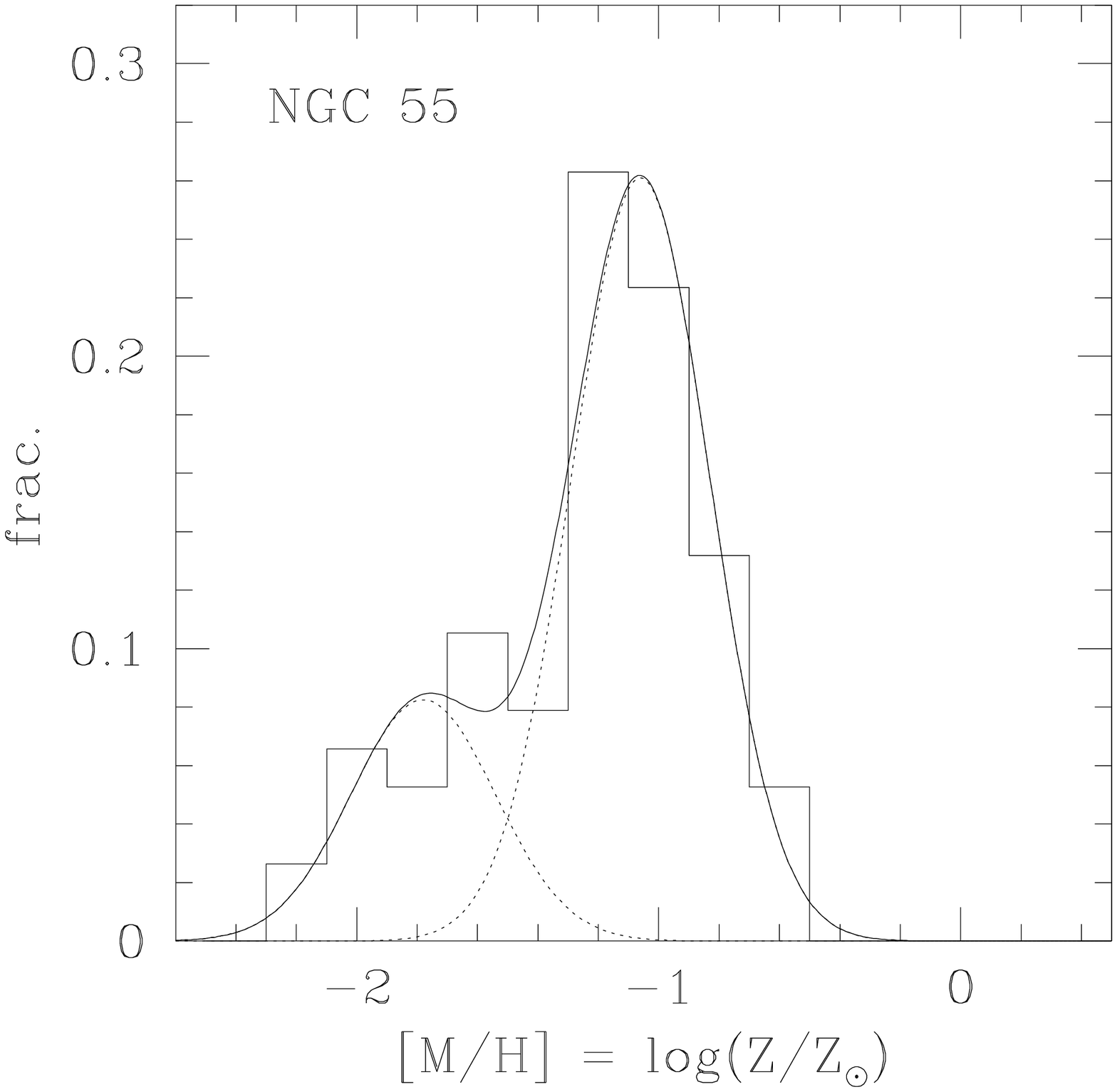}
\includegraphics[height=2.in,width=2.in]{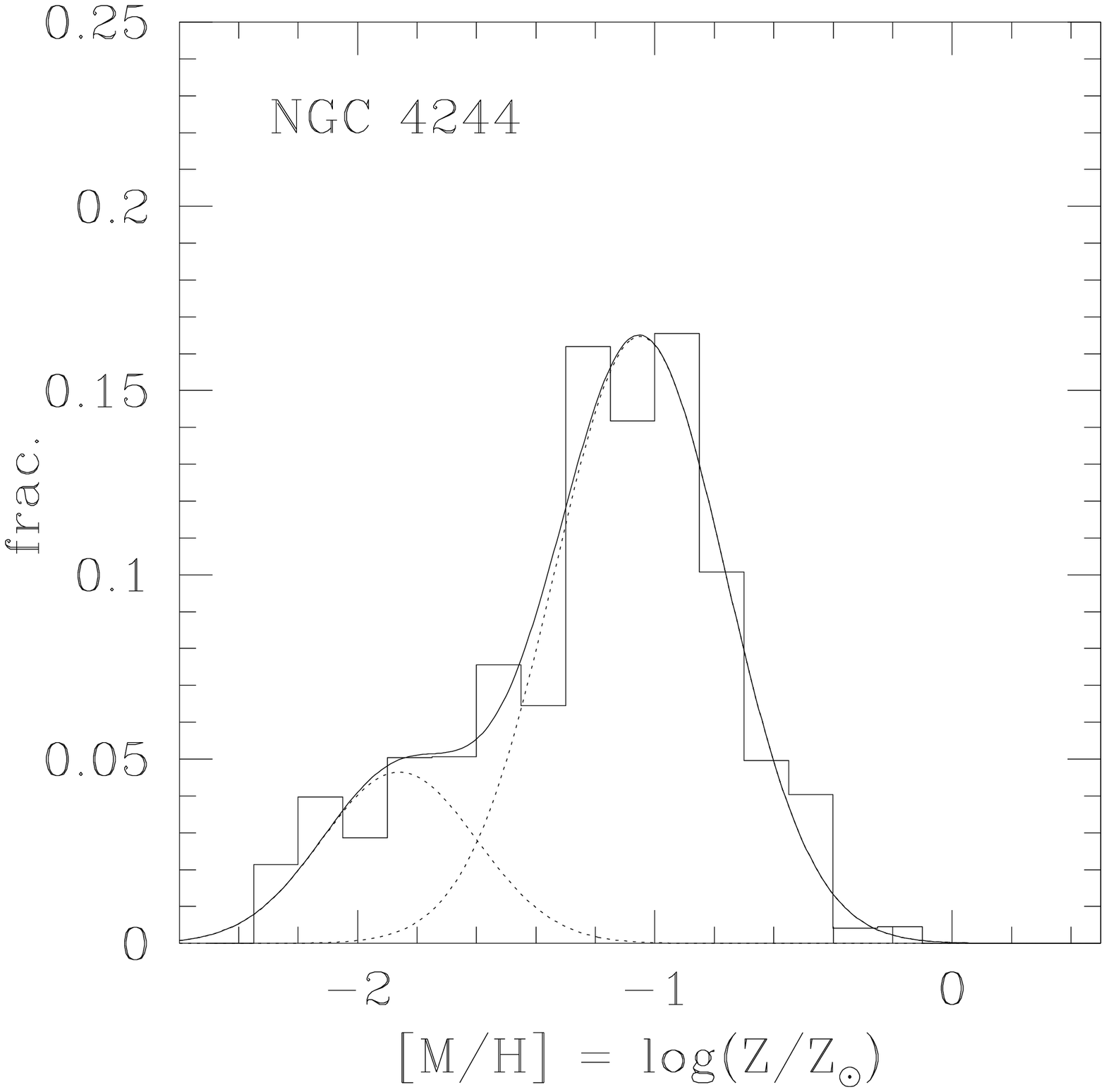}
\includegraphics[height=2.in,width=2.in]{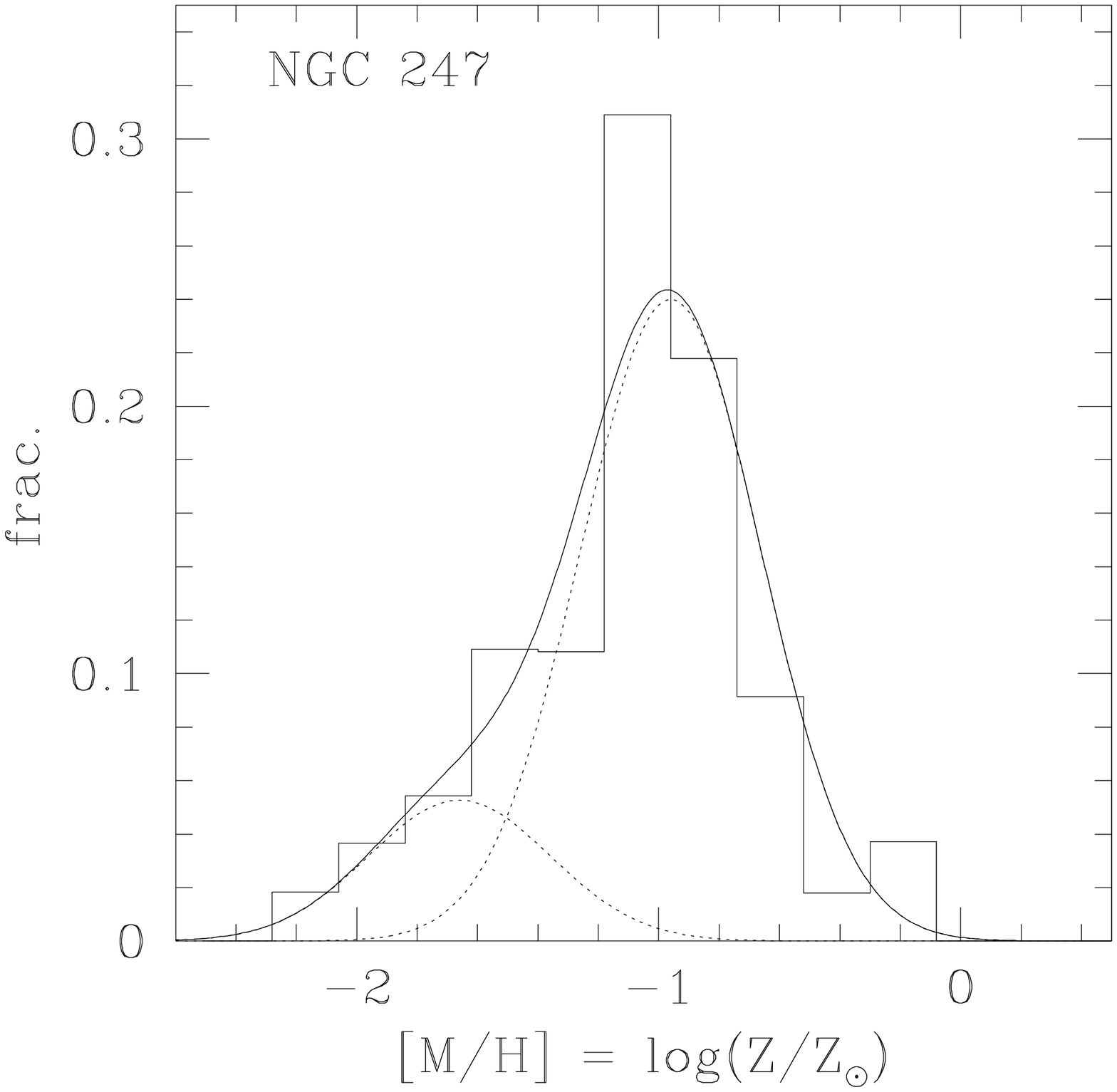}
\includegraphics[height=2.in,width=2.in]{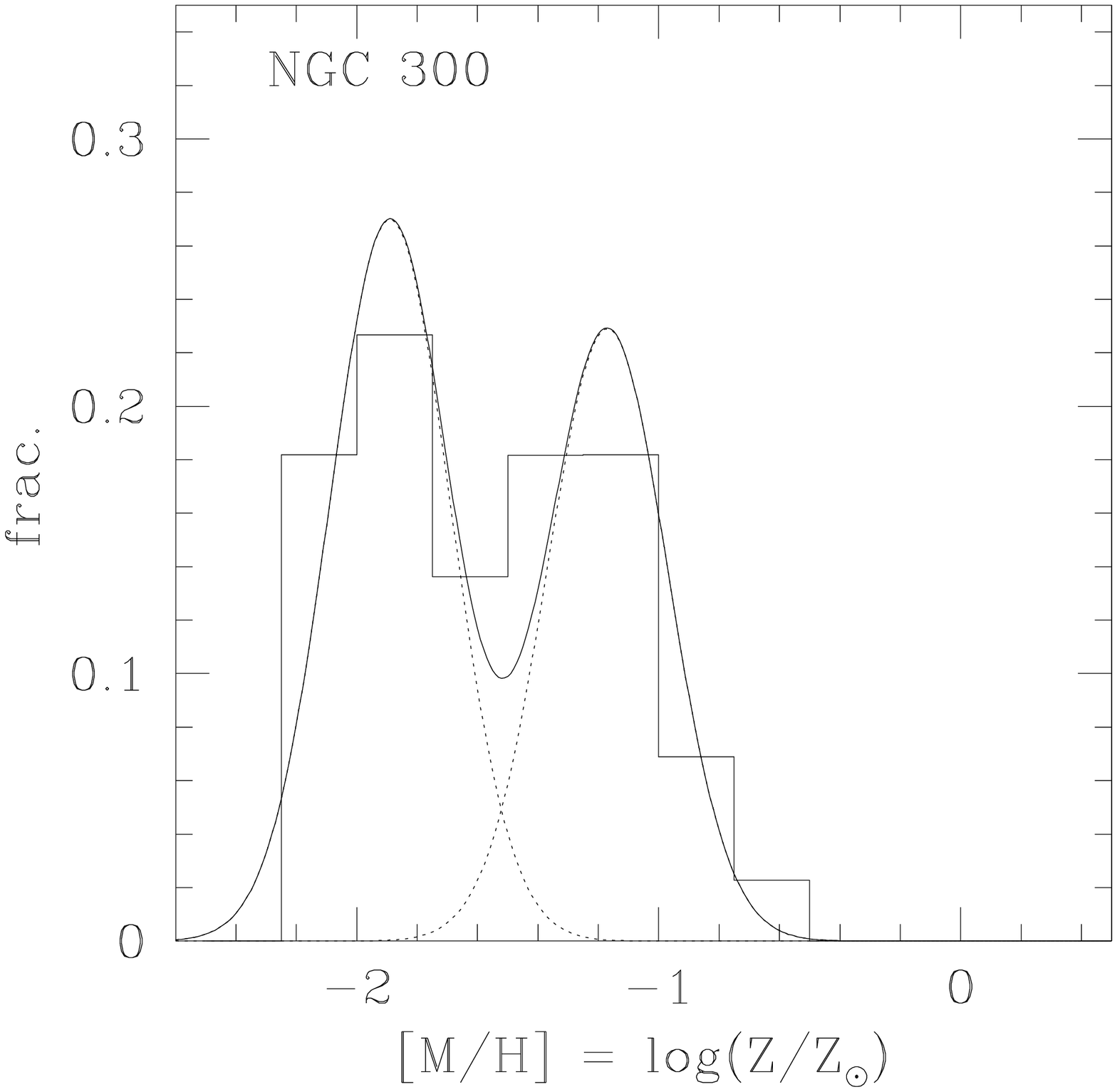}
\caption{Metallicity distribution functions for the sample galaxies, 
with best fitting two Gaussian (dotted lines), and the combination 
of both. Each distribution has been normalized by the total number 
of stars. The histograms and the mixture models are plotted such 
that they both enclose the same area.}
\label{gauss2}
\end{figure*}

The histograms shown in Fig.\,\ref{gauss2} represent the measured 
incompleteness-corrected metallicity distribution functions for 
all galaxies in the sample, ordered by decreasing luminosity. 
An optimal bin width was estimated on a case-by-case basis 
according to the star sample size and the distribution's skewness 
(Scott 1992). In each metallicity distribution a strong relatively
metal-rich peak is prominent, and each distribution is asymmetric 
about this peak. The issue is to determine how strongly a two
Gaussian fit is preferred at any level in the metallicity 
distribution functions over simpler models such as a Gaussian 
distribution. The properties of metallicity distribution are 
quantified using the widely used Kaye's Mixture Model (KMM) 
statistical test (Ashman, Bird, \& Zepf 1994 and references 
therein). The KMM test uses the maximum likelihood technique to 
test if a distribution is better modelled as a sum of two Gaussian 
than as a single Gaussian (the null hypothesis). The $p$-value 
returned by KMM adequately measures the statistical significance 
of the improvement in the fit in going from one to two groups. 
As there is no reason to assume that both populations in the fit 
have similar dispersions, a mixture model fitting is applied to 
the data such that the dispersion for the two populations are not 
constrained to be identical, i.e., the so-called heteroscedastic 
fitting. 

Ashman et al. (1994) have cautioned that the output likelihood 
in the case of heteroscedastic fitting is difficult to interpret, 
and propose to perform bootstrap simulations to assess the result 
of the fitting. We generate a large number of synthetic samples 
by randomly selecting stars with replacement from the original 
data set, constructing the metallicity distribution, and finally 
applying the KMM test to the simulated samples. The distributions 
of the mixture fit parameters for the synthetic samples are 
fitted by Gaussian, and then their means and dispersions are 
taken to be the final estimates of the mean value and the formal 
error for each fit parameter. 

The parameters for the best fit models for the two peaks, referred 
to as the metal-poor and the metal-rich hereafter, as assigned by 
the KMM test are presented in Table \ref{fit_gauss}.  
Fig.\,\ref{gauss2} shows the double Gaussian distributions 
corresponding to the metallicity peaks and their sum. The $p$-value 
is close to zero in all cases, indicating high probabilities that 
two Gaussian are better fits to the metallicity distribution 
functions than single ones. The uncertainties in distance modulus 
and the foreground extinction have rather small impact in all 
fitted parameters: less than 0.1-0.15 dex for the mean metallicities, 
0.05 dex for the dispersion, and a few percent for the fraction of 
stars in each stellar component, with the overall shape of the 
metallicity distribution unchanged.   

Using the mixture model estimates, we can investigate the behavior
of metal-rich and metal-poor stellar populations as a function of 
host galaxy luminosity. An obvious question is whether a stellar 
metallicity vs. galaxy luminosity relation and/or other correlations 
are present for both subcomponents, as for both blue/metal-poor 
and red/metal-rich globular clusters of early-type galaxies. 
An important question is to know whether the stellar 
luminosity-metallicity relation reported by Mouhcine et al. (2005a) 
might result from different mixture of two populations with roughly 
constant metallicities. We have augmented our database with similar 
published measurements for the stellar populations in the outskirts 
of M~31 (Durrell et al. 2001), and the giant E/S0 NGC~5128 (Harris 
\& Harris 2000).

\begin{figure*}
\includegraphics[height=3.2in]{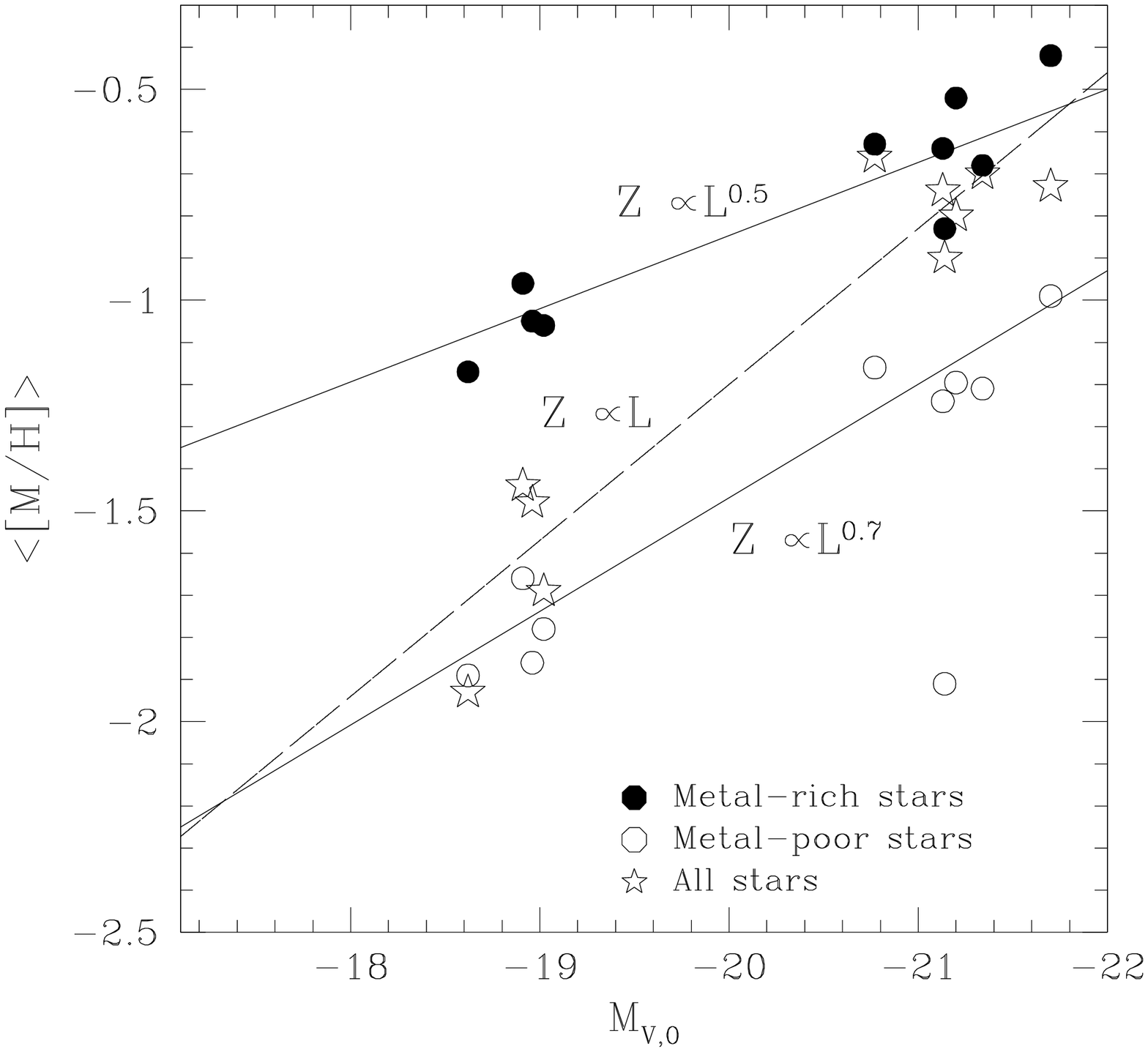}
\includegraphics[height=3.2in]{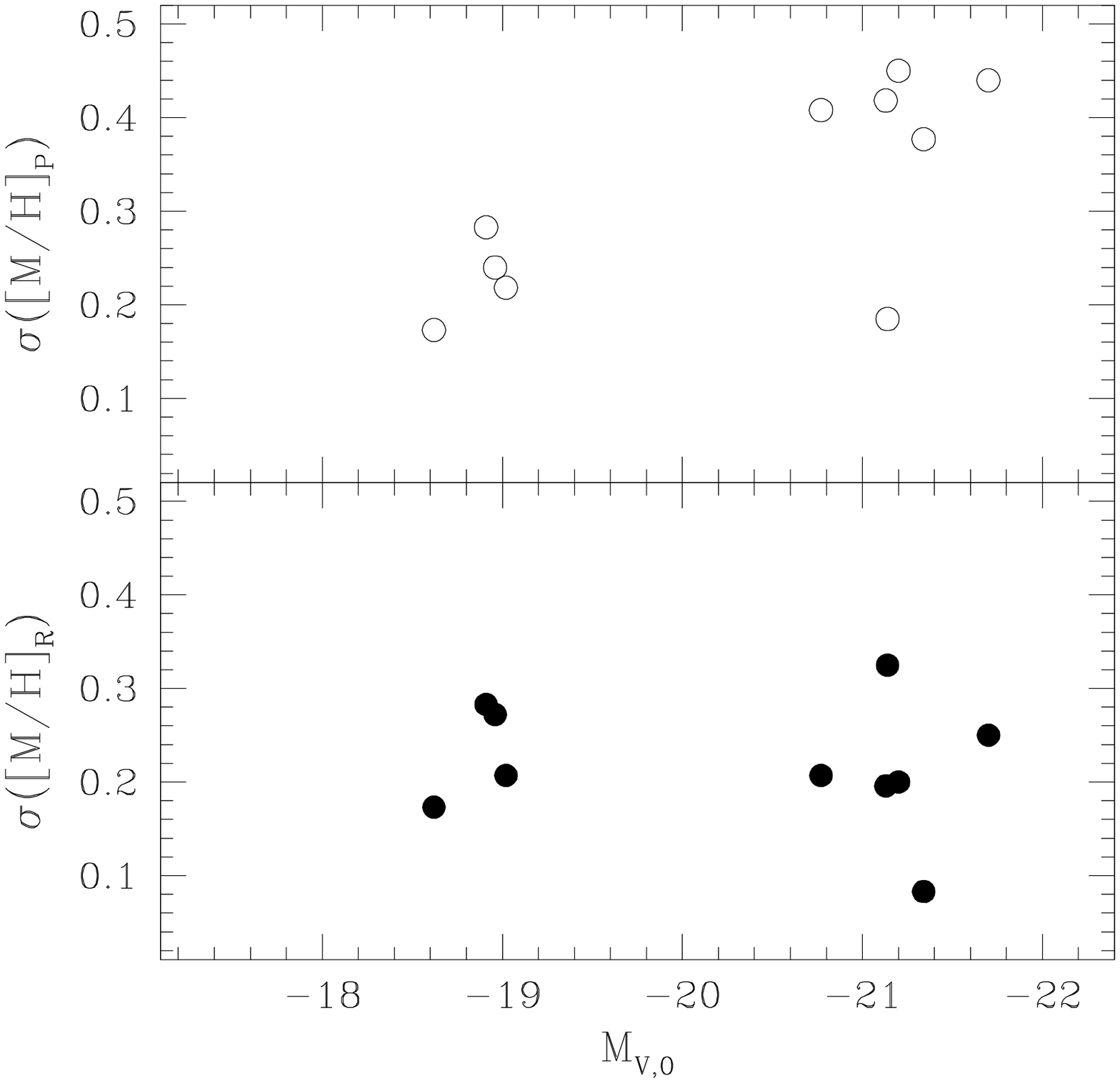}
\caption{Left: The relationship between the mean metallicities 
of the metallicity distribution peaks as found by the KMM test, 
and the host galaxy luminosity. The filled dots show the mean 
metallicity of the metal-rich population, the circles show the 
mean metallicity of the metal-poor population. Open stars show 
the relationship between the mean stellar metallicity and the 
host galaxy luminosty. Right: Relationships between the 
metallicity dispersion of the metal-rich (bottom) and the 
metal-poor (top) stellar subpopulations and the host galaxy 
luminosity. }
\label{zl_subpop}
\end{figure*}

The left panel of Fig.~\ref{zl_subpop} shows the relationships 
between mean metallicities of the stellar population subcomponents 
as returned by the KMM test, the mean metallicity of field stars 
as derived by Mouhcine et al. (2005b), and the host galaxy V-band 
luminosity. The metallicities of both metal-poor and metal-rich 
stellar population subcomponents in the outskirts of spirals 
clearly increase for brighter host galaxies. The slopes of the 
luminosity-metallicity correlations are similar for both stellar 
subcomponents. A consequence of this is a nearly constant offset 
between the metallicities of the two subcomponents across nearly 
three orders of magnitude in galaxy luminosity. 
The observed similarity suggests that the conditions of metal-rich 
and metal-poor star formation could not have been too different 
across a large range of host galaxy luminosity/mass. 
The metallicity-luminosity relation for the total resolved stellar 
populations is much steeper than those seen for both subpopulations. 
This suggests that the increasing metallicity of the stellar 
populations in the outer regions for brighter host galaxies is due 
to an increasing contribution of metal-rich stars. 

The right panel of Fig.~\ref{zl_subpop} shows the variation of 
the fitted metallicity dispersion for the stellar subpopulations
as a function of parent galaxy luminosity. The metallicity 
dispersion of metal-poor subpopulations correlates with galaxy 
luminosity, i.e., brighter galaxies tend to have metal-poor 
subpopulations with larger metallicity dispersions than faint 
galaxies. However, the metallicity dispersions of metal-rich 
subpopulations lack any correlation with galaxy luminosity, and 
scatter around the mean metallicity dispersion of $\sim 0.25$, 
smaller than the metallicity dispersion of metal-poor stellar
subpopulations. 

The properties of the stellar populations in the outskirts of 
M~31, and more interestingly NGC~5128, agree nicely with the 
correlations defined by our galaxy sample. This suggests that 
the formation of field stars in the outer regions of early-type 
galaxies might share similarities with those in the outer
regions of spirals, in agreement with the conclusion of Forbes 
et al. (2001) regarding globular cluster systems. 
This supports the idea that the diffuse stellar populations in 
the outer regions of galaxies exhibit a dual nature reflecting 
the presence of two major stellar populations that share the 
properties of each globular cluster subpopulations (see Forte 
et al. 2005). 

\section{Implications for the formation of the outskirts of
spirals}
\label{impli}

We have thus found that the properties of field stars in the 
outer regions of spirals are correlated with parent galaxy 
luminosity. Given these correlations, what may be the nature 
of the diffuse stellar populations in the outskirts of spiral
galaxies? The correlations of mean metallicities of both 
subcomponents with galaxy luminosity indicate that the 
formation of both metal-rich and metal-poor stellar populations 
is linked to their host galaxies.

These constraints fit well into the {\it{in situ}} scenario 
in which at least a significant fraction of field stars in the 
outer regions of spiral galaxies has been formed after the 
parent galaxies assembled into individual entities. The similar 
range of metallicities covered by metal-poor stars and the 
Local Group dwarf galaxies indicates that metal-poor stellar 
populations may be built up by the disruption of protogalactic 
fragments with sizes similar to those of the present-day 
Local Group dwarf galaxies. This does not imply however that 
the disrupted galactic fragments have stellar formation 
histories similar to those of dwarf galaxies in the Local 
Group (see e.g., Shetrone et al. 2003, Venn et al. 2004 for 
a similar conclusion regarding the formation of the Galaxy 
stellar halo). The chemical properties of stars in the outer 
regions of spirals beyond the Local Group are not available 
yet to constrain the details of their star formation histories. 
The correlation of the mean metallicities of metal-poor field 
stars with the host galaxy luminosity suggests that the 
disrupted fragments were already embedded in the dark matter 
halo of the final galaxy after its assembly rather than being 
independent satellites. 

The mean metallicities of metal-rich subcomponents suggest 
that they formed in protogalactic fragments that are more 
massive, and thus more evolved chemically, than progenitors of 
present-day dwarf galaxies in the Local Group. The correlation 
of mean metallicities of metal-rich components with galaxy 
luminosity indicates that metal-rich stars knew about the size 
of the final galaxy to which they belong. 

A metallicity dispersion could be the result of star formation 
events where stars form over a period of time long enough to 
produce different generations of stars with different chemical
abundances, or also by the merging of different stellar 
systems that were chemically isolated from each other, whether 
spatially or temporally (see Peng et al. 2006). The latter 
scenario could apply to the metal-poor component in the 
outskirts of spirals formed from the disruption of isolated 
fragments. The metallicity dispersion of dwarfs in the Local 
Group does not depend on galaxy luminosity (see C\^ot\'e et al. 
2000 for more details). The metallicity dispersion of a 
disrupted population of fragments with sizes similar to those 
of present-day dwarfs is expected to increase with the host 
galaxy luminosity/mass.
The narrower metallicity dispersion of metal-rich components, 
especially for bright galaxies, indicates that the metal-rich 
stars may be formed in few large star formation events where 
the gas is well mixed. The lack of a correlation between the 
metallicity dispersion of metal-rich subcomponents and galaxy 
luminosity suggests that these star formation events were 
possibly uncorrelated with the assembly of galactic disks, 
which dominate the light budget in present-day intermediate 
and late-type spiral galaxies. 

One of the most pressing questions regarding spiral galaxy 
evolution is the connection between globular cluster systems 
and field stars. The disruption of globular clusters, revealed 
by tidal tails (e.g., Grillmair et al. 1995), may contribute 
to the formation of the stellar component at the outskirts of 
galaxies (Aguilar et al. 1988; see Freeman \& Bland-Hawthorn 
2002 for a detailed discussion). 
Our knowledge of globular cluster systems in spiral galaxies 
is still limited to a handful of galaxies (e.g., Harris 1991; 
Ashman \& Zepf 1998; Rhode \& Zepf 2003). A view is emerging 
however that inner metal-rich globular clusters in spiral 
galaxies may be associated with their bulges (Minnitti 1995; 
C\^ot\'e 1999; Barmby et al. 2001; Perrett et al. 2002; 
Forbes et al. 2001; Goudfrooij et al. 2003). Stars belonging 
to the metal-rich component could be identified as field 
stars associated with the formation of metal-rich globular 
clusters. If this is indeed how the metal-rich components 
form, one would expect their mean metallicities to vary 
systematically along the Hubble sequence of spirals, and 
metal-rich stars in the outskirts of spirals to show 
kinematics similar to bulge stars. The strong rotation of 
both globular clusters and field stars in the inner speroid 
of M~31 (Perrett et al. 2002; Hurley-Keller et al. 2004), 
as well as in the outer halos of some giant ellipticals 
(C\^ot\'e et al. 2003, Peng et al. 2004) support the 
existence of a connection between metal-rich field stars 
and metal-rich globular globular clusters. For M~31, Ibata 
et al. (2005) have argued however that the rotating structure 
is related to an extended disk rather than to the bulge. 

The formation scenario of field star populations in the 
outskirts of spirals discussed here shares similarities 
with the scenarii proposed by Forbes et al. (1997) and 
Rhode \& Zepf (2004) to account for the observed properties 
of globular cluster systems of early-type galaxies. 
Peng et al. (2006) have conducted a homogeneous study of 
globular cluster systems of a large sample of early-type 
galaxies in the Virgo cluster. The observed overall behavior 
of globular cluster systems, i.e., the correlation of both 
subpopulation mean metallicities with galaxy luminosity, 
the correlation of metallicity dispersion of metal-poor 
globular clusters with galaxy luminosity, and the lack of 
a correlation between galaxy luminosity and metallicity 
dispersion of metal-rich globular clusters, are similar 
to what is established here for field stars in the outskirts 
of spirals. This suggests that the formation histories of 
the stellar populations in the outer regions of both 
early-type galaxies and spirals might have features in common. 
However, the details of the scaling relations are different. 
The subcomponents of globular cluster systems of early-type 
galaxies show shallower luminosity-metallicity relations, 
and larger metallicity dispersions than those of field stars
populations in the outskirts of spirals. 
The implications of those differences on the understanding 
of the assembly histories of stellar populations in the outer 
regions of galaxies are not clear due to the small number of 
galaxies in our sample, which affect the determination of the 
slopes of the luminosity-metallicity relations of field star 
subcomponents, and the uncertainties affecting the 
determination of the absolute globular cluster metallicities 
from braod band colors (see Peng et al. 2006 for a detailed 
discussion of this issue). 

The formation scenario of the outer regions of spiral galaxies
that emerges here is, as yet, only an approximative (and 
speculative) outline. Given the importance of this issue in 
the context of our understanding of the formation of galaxies, 
more data are needed to investigate the variation of chemical 
abundances, kinematical properties, and spatial distributions
of field stars in the outer regions of galaxies over large 
fields, dependence on galaxy morphological type etc, in order 
to test the observational consequences of such a scenario.

\section*{Acknowledgements}
I would like to thank warmly W. E. Harris, D. Geisler, S. van 
den Bergh for their critical reading of earlier versions of the 
paper, B.K. Gibson and R. Ibata for enlightening discussions, 
and the anonymous referee for useful comments which significantly 
improved the paper.

\begin{table}
\caption{Results of the KMM test applied to the metallicity distribution
of the sample galaxies. Columns: (1) Galaxy name; (2) galaxy magnitude;
(3) the statistical significance that the MDF is unimodal;
(4) fraction of the metal-rich peak; (5) metallicity of the metal-rich
peak; (6) spread of the metal-rich peak; (7) fraction of the metal-poor
peak; (8) metallicity of the metal-poor peak; (9) spread of the metal-poor
peak. }
\begin{tabular}{lrrrrrrrr}
\hline
Galaxy & $M_{V,\circ}$ & $p$ &$f_R$ & $[Fe/H]_R$ & $\sigma_R$ & $f_P$ &  $[Fe/H]_P$ &
$\sigma_P$ \cr
\hline
NGC~300  & -18.62 & 0.004 & 0.459 & -1.17 & 0.20 & 0.541 & -1.89 & 0.20   \cr
NGC~247  & -18.91 & 0.051 & 0.820 & -0.96 & 0.30 & 0.180 & -1.66 & 0.30   \cr
NGC~4244 & -18.96 & 0.000 & 0.798 & -1.05 & 0.29 & 0.202 & -1.86 & 0.26   \cr
NGC~55   & -19.02 & 0.006 & 0.752 & -1.06 & 0.23 & 0.248 & -1.78 & 0.24   \cr
NGC~4945 & -20.77 & 0.000 & 0.799 & -0.63 & 0.23 & 0.201 & -1.16 & 0.42   \cr
NGC~253  & -21.13 & 0.000 & 0.616 & -0.64 & 0.22 & 0.384 & -1.24 & 0.43   \cr
NGC~3031 & -21.14 & 0.000 & 0.884 & -0.83 & 0.34 & 0.116 & -1.91 & 0.21   \cr
NGC~4258 & -21.34 & 0.000 & 0.476 & -0.68 & 0.13 & 0.524 & -1.21 & 0.39   \cr
\hline
\end{tabular}

\label{fit_gauss}
\end{table}

\end{document}